\documentclass[aps,prl,twocolumn,showpacs,superscriptaddress,amsmath,amssymb,floatfix]{revtex4-1}
\usepackage{graphicx}
\usepackage{dcolumn}
\usepackage{bm}
\begin{document}

\title{Liquid-Gas Asymmetry and the Wavevector-Dependent Surface Tension}

\author{A.O.\ Parry}
\affiliation{Department of Mathematics, Imperial College London, London SW7 2BZ, UK}

\author{C.\ Rasc\'{o}n}
\affiliation{GISC, Departamento de Matem\'aticas, Universidad Carlos III de Madrid, 28911 Legan\'es, Madrid, Spain}

\author{R.\ Evans}
\affiliation{HH Wills Physics Laboratory, University of Bristol, Bristol BS8 1TL, United Kingdom }

\begin{abstract}
Attempts to extend the capillary-wave theory of fluid interfacial fluctuations to microscopic wavelengths, by introducing an effective wave-vector ($q$) dependent surface tension $\sigma_\textup{eff}(q)$, have encountered difficulties. There is no consensus as to even the shape of $\sigma_\textup{eff}(q)$. By analysing a simple density functional model of the liquid-gas interface, we identify different schemes for separating microscopic observables into background and interfacial contributions. In order for the backgrounds of the density-density correlation function and local structure factor to have a consistent and physically meaningful interpretation in terms of weighted bulk gas and liquid contributions, the background of the total structure factor must be characterised by a microscopic $q$-dependent length $\zeta(q)$ not identified previously. The necessity of including the $q$ dependence of $\zeta(q)$ is illustrated explicitly in our model and has wider implications, i.e. in typical experimental and simulation studies, an indeterminacy in $\zeta(q)$ will always be present, reminiscent of the cut-off used in capillary-wave theory. This leads inevitably to a large uncertainty in the $q$ dependence of $\sigma_\textup{eff}(q)$.
\end{abstract}



\maketitle

Understanding the nature of the interface separating coexisting fluid phases has provided fundamental insights into the properties of matter, including the necessity for attractive intermolecular forces. Of particular importance is the role played by thermally induced interfacial wandering. Classical Capillary-Wave (CW) theory \cite{BLS1965,Wertheim1976,Weeks1977,Evans1979,Rowlinson1982,Evans1990,Aarts2004} provides a remarkably successful description of the fluctuation properties of fluid interfaces, leading directly to the concepts of interfacial roughness and entropic repulsion. These underpin our modern understanding of interfacial phase transitions. Recent extensions of CW theory that include a wavevector dependent tension attempt to describe interfacial fluctuations at microscopic scales. However, these have proved problematic \cite{Rochin1991,Blokhuis1993,Napiorkowski1993,Parry1995,Mecke1999,Blokhuis1999,Fradin2000,Mora2003, Tarazona2012,Blokhuis2008,Blokhuis2009,Hofling2015,Parry2014}. For example, there is no consensus as to the \textit{sign} of the rigidity coefficient characterising the proposed order $q^2$ correction to the equilibrium surface tension. In the present paper, we provide a worked example, using a simple analytically solvable density functional theory (DFT), which illustrates that problems have arisen because previous analyses have failed to identify a $q$ dependent microscopic lengthscale $\zeta(q)$ crucial for any treatment attempting to extend CW theory to shorter wavelengths.\\

The central idea of CW theory is that the thermal excitations of long-wavelength undulations in the local height $\ell(\bf{x})$ of a liquid-gas interface are resisted by the equilibrium surface tension $\sigma$ . Thus, in the absence of additional pinning effects, such as gravity, the thermal average of the fluctuations satisfies $\langle\,|\tilde\ell({\bf{q})}|^2\,\rangle={k_B T}/\sigma q^2$, where $\tilde\ell({\bf{q}})$ are the Fourier amplitudes of $\ell(\bf{x})$, with ${\bf x}=(x,y)$. This result is valid only for $q\ll\Lambda$, where $\Lambda$ is a cut-off of order the inverse bulk correlation length. The small $q$ singularity is also manifest in microscopic observables such as the parallel Fourier transform of the density-density correlation function $G(z,z';q)$ and local structure factor $S(z;q)\!=\!\int\! dz'\, G(z,z';q)$. Since, in the long-wavelength limit, fluctuations of $\ell(\bf{x})$ translate the density profile, CW theory implies that, in the interfacial region, 
\begin{equation}
G(z,z';q)\approx\frac{\,\rho'(z)\rho'(z')}{\sigma q^2},\hspace{0.5cm} S(z;q)\approx\frac{\,\Delta\rho\,\rho'(z)}{\sigma q^2}
\label{CW2}
\end{equation}
for $q\ll\Lambda$. Here, $\rho(z)$ is the equilibrium density profile, $\Delta\rho=\rho_l\!-\!\rho_g$ is the difference in bulk coexisting densities, and we have set $k_B T=1$. Hence, for $q\ll\Lambda$,  we might anticipate that the total structure factor $S(q)\!=\!\int_{-L_g}^{L_l}\!\! dz\; S(z;q)$, behaves approximately as
\begin{equation}
S(q)\;\approx\;\, L_gS_g+L_lS_l\;+\;\frac{(\Delta\rho)^2}{\sigma q^2}
\label{CW3}
\end{equation}
where $L_g$ and $L_l$ are the macroscopic sizes of the gas and liquid phases, which have ($q$-dependent) bulk structure factors $S_g$ and $S_l$, respectively. In (\ref{CW3}), the first two terms are the expected "background" arising from the bulk phases and the third is the "excess", containing the Goldstone mode contribution.\\

It is certainly natural to ask if the success of CW theory is limited only to long-wavelengths ($q\ll\Lambda$). A particular issue, hotly debated, is whether one can extend the theory to allow for a wavevector-dependent surface tension $\sigma_\textup{eff}(q)$ that replaces $\sigma$ in (\ref{CW2}) and (\ref{CW3}) and then meaningfully apply this interfacial based description at the microscopic scale. For example, it is accepted that dispersion forces lead to a non-analytic term in the low $q$ expansion that can be assimilated as $\sigma_\textup{eff}(q)=\sigma+ Aq^2 \ln q+\cdots$  \cite{Mecke1999,Blokhuis1999,Parry2014}, similar to curvature corrections to the tension of a spherical wall-fluid interface \cite{*[{}][{, and references therein.}] Stewart2005}. Here, the coefficient $A>0$ is proportional to the coefficient of $r^{-6}$ in the interatomic pair potential, and is insensitive to the precise definition of the interface location. A more contentious question is whether or not one can identify a physically meaningful $\sigma_\textup{eff}(q)$ over the whole range of wavevectors reaching to the inverse atomic scale. Two strategies are possible. The original suggestion was to identify $\ell({\bf{x}})$ from a given molecular configuration and then define a $q$-dependent tension from $\langle|\tilde\ell({\bf{q})}|^2\rangle$ \cite{Mecke1999}. For this, one may now use sophisticated many-body definitions of the interface which go far beyond the notion of a local Gibbs dividing surface \cite{Chacon2003,Tarazona2007,Tarazona2012,Fernandez2013}. However, one cannot then infer directly the behaviour of the correlation function and structure factors because, away from the $q\ll\Lambda$ limit, interfacial fluctuations no longer merely translate the profile \cite{Tarazona2012,Parry2014,Fernandez2013}. Moreover, one does not know what are the "bulk" contributions to measured observables \cite{Tarazona2012}. A different but related strategy, linked more directly to scattering experiments, is to start from the microscopic observables, such as the measured total structure factor $S(q)$, and identify an effective tension from the excess contribution \cite{Blokhuis2008,Blokhuis2009,Hofling2015}, i.e.~one defines $\sigma_\textup{eff}(q)$ via:
\begin{equation}\label{sigmaeff}
S^{ex}(q)\;\equiv\;\frac{(\Delta\rho)^2}{\,\sigma_\textup{eff}(q)\, q^2}\;.
\end{equation}
This approach has the advantage that $\sigma_\textup{eff}(q)$ does not depend on the definition of the interface position. However, it still requires that we first write $S(q)\;=\;S^{bg}(q)+S^{ex}(q)$, and determine a suitable background $S^{bg}(q)$ in terms of $S_g$ and $S_l$. In this paper, we use a simple density functional theory to illustrate that, if such a separation is imposed, one is required to write
\begin{equation}
 S^{bg}(q)\,=\,\big(L_g+\zeta(q)\big)\,S_g\,+\,\big(L_l-\zeta(q)\big)\,S_l
\label{Sback}
\end{equation}
where $\zeta(q)$ is a microscopic \textit{$q$-dependent} lengthscale which weights the bulk gas and liquid contributions. The properties of $\zeta(q)$ and the limitations placed on the accessibility of $\sigma_\textup{eff}(q)$ from measured structure factors are described. We argue that, in general, an uncertainty in  $\sigma_\textup{eff}(q)$ will always arise from that in $\zeta(q)$, implying that in experiments, simulations, and more realistic DFTs, it will be extremely difficult to identify a robust $q$-dependent tension from measurements of only $\rho(z)$, and $S(z;q)$ or $S(q)$, beyond perhaps the leading-order terms of its small $q$ expansion.\\

We consider a square-gradient description of the interface based on the Grand Potential functional \cite{Evans1979}
\begin{equation}
\Omega[\rho]=\int\!\!d{\bf r}\; \left\{\;\frac{b}{2}\, (\nabla \rho)^ 2\, +\, \phi(\rho) \,\right\}
\label{LGW}
\end{equation}
The free-energy density $\phi(\rho)$ is a double-well potential modeling the coexistence of gas and liquid phases below the critical temperature which have different bulk structure factors $S_g={1}/{b(\kappa_g^2\!+\!q^2)}$ and $S_l=1/{b(\kappa_l^2\!+\!q^2)}$. Here, $\kappa_g\equiv\sqrt{\phi''(\rho_g)/b}$ and $\kappa_l\equiv\sqrt{\phi''(\rho_l)/b}$ identify the bulk correlation lengths $\xi_g=1/\kappa_g$ and $\xi_l=1/\kappa_l$, which are implicitly temperature dependent. Minimization of $\Omega[\rho]$ leads to $b\rho''(z)=\phi'(\rho)$, the solution of which determines the free interfacial profile over the macroscopic interval $[-L_g,L_l]$. It is well known that the square-gradient theory does not incorporate the CW broadening of the density profile. However, the identification (\ref{sigmaeff}) only involves the difference in the bulk densities, and should be determined reliably even at mean-field level. The local structure factor then follows from the Ornstein-Zernike equation
\begin{equation}
b\left(-\frac{{\partial}^2}{\partial z^2}+q^ 2 +\frac{1}{b}\,\phi''(\rho(z))\right)S(z;q)=1
\label{OZ}
\end{equation}
We now consider a Double Parabola (DP) potential $\phi(\rho)=b\kappa_g^2(\rho-\rho_g)^2/2$ for $\rho<\rho_0$ and $\phi(\rho)=b\kappa_l^2(\rho-\rho_l)^2/2$ for $\rho>\rho_0$, where $\rho_0=f\rho_g+(1-f)\rho_l$ and $f=\kappa_g/(\kappa_g+\kappa_l)$. The profile follows as
\begin{equation}\label{Profile}
\rho(z)=
\begin{cases}
\,\rho_g+(\rho_0-\rho_g)e^{-\kappa_g|z|} & \text{if }\, z<0\\[.15cm]
\;\rho_l+(\rho_0-\rho_l)e^{-\kappa_l z} & \text{if }\, z>0
\end{cases}
\end{equation}
Here, we have located the origin such that $\rho(0)=\rho_0$, where the gradient $\rho'(0)=\Delta\rho/(\xi_g+\xi_l)$ is largest. A straightforward calculation gives $\sigma=b(\Delta\rho)^2/2(\xi_g+\xi_l)$. Henceforth, we set $b=1$. The liquid/gas asymmetry means that the Gibbs dividing surface $Z_G$, defined by
\begin{equation}\label{ZG}
\int\! dz\,\Big(\,\rho(z)-\rho_g-(\rho_l-\rho_g)\,\theta\big(z-Z_G\big)\Big)=0\,,      
\end{equation}
lies at $Z_G=\xi_l\!-\!\xi_g$, away from the origin (see \cite{SM}). $\theta(z)$ is the Heaviside step function. The local structure factor follows from Eq.~(\ref{OZ}):
\begin{equation}\label{Szq}
\hspace*{-.15cm}S(z;q)=
\begin{cases}
\,S_g+(S(0;q)-S_g)\,e^{-\kappa_g(q)|z|} & \text{if }\, z<0\\[.15cm]
\,S_l+(S(0;q)-S_l)\,e^{-\kappa_l(q) z} & \text{if }\, z>0
\end{cases}
\end{equation}
and attains its maximum value at the origin
\begin{equation}
S(0;q)\;=\;\frac{1}{q^2}\,\frac{\kappa_g(q)\,S_g+\kappa_l(q)\,S_l}{\,(\kappa_g(q)+\kappa_g)^{-1}+(\kappa_l(q)+\kappa_l)^{-1}}
\end{equation}
where $\kappa_g(q)^2\equiv\kappa_g^2+q^2$, and $\kappa_l(q)^2\equiv\kappa_l^2+q^2$. One may also determine $G(z,z';q)$ \cite{SM}; for example, $G(0,0;q)=S(0;q)/\big(\kappa_g(q)S_g+\kappa_l(q)S_l\big)$. Note that, in contrast to the density profile, the decay of $S(z;q)$ depends on $q$, meaning that the expression for $S(z;q)$ is consistent with the CW expectation (\ref{CW2}) only at small $q$, where $\kappa_g(q)\approx\kappa_g$ and $\kappa_l(q)\approx\kappa_l$. Consequently, no single $q$-dependent surface tension $\sigma_\textup{eff}(q)$ can replace $\sigma$ in all of the expressions (\ref{CW2})-(\ref{CW3}).\\

We wish to repackage our results as $G(z,z';q)=G^{bg}(z,z';q)+G^{ex}(z,z';q)$ {\it and} $S(z;q)=S^{bg}(z;q)+S^{ex}(z;q)$, where the separation is done consistently (for example, $S^{bg}(z;q)=\int dz' G^{bg}(z,z';q)$), and is subject to a few necessary reasonable physical constraints \cite{SM}. Specifically, $G^{bg}(z,z';q)$ must be continuous, contain lengthscales only defined for the bulk fluids and decay to the appropriate liquid and gas correlation functions when $z$ and $z'$ are far from the interfacial region. These requirements ensure that $S^{bg}(z;q)$  is continuous which is also necessarry since, otherwise, $S^{ex}(z;q)$ would be discontinuous, which would not match sensibly with (\ref{CW2}). For example, it is not appropriate to suppose $S^{bg}(z;q)$ jumps from $S_g$ to $S_l$ at $z=Z_G$, say, since $S^{ex}(z;q)$ would be inconsistent with CW theory even as $q\to 0$. This means that, within our model, the {\it only} allowed separation of the local structure factor, which is physically meaningful, is of the form
\begin{equation}\label{Szqback}
S^{bg}(z;q)=
\begin{cases}
\,S_g+(S^{bg}(0;q)-S_g)\,e^{-\kappa_g(q)|z|} & \text{if }\, z<0\\[.15cm]
\,S_l+(S^{bg}(0;q)-S_l)\,e^{-\kappa_l(q) z} & \text{if }\, z>0
\end{cases}
\end{equation}
and $S^{ex}(z;q)=S^{ex}(0;q)\,e^{-\kappa_g(q)|z|}$ for $z<0$, and similarly for $z>0$ but with $\kappa_l(q)$. This way, the background contribution varies continuously from $S_g$ to $S_l$ through the interface. Here, $S(0;q)=S^{bg}(0;q)+S^{ex}(0;q)$  with consistency demanding that  $S^{bg}(0;q)=G^{bg}(0,0;q)(\kappa_g(q)^{-1}+\kappa_l(q)^{-1})$ \cite{SM}. We note that, without loss of generality, one can always write
\begin{equation}\label{fractionG}
G^{bg}(0,0;q)\,=\,f_\textsc{G}(q)\, G_g+(1-f_\textsc{G}(q))\,G_l
\end{equation}
where $G_g=1/2\kappa_g(q)$, $G_l=1/2\kappa_l(q)$ are the 2D Fourier transforms of the bulk gas and liquid correlation functions \cite{Parry2014} and $f_\textsc{G}(q)$ is the "fraction" of the bulk gas contributing to the background correlation function at the $z=0$ plane. Similarly, we can always write 
\begin{equation}
S^{bg}(0;q)\,=\,f_\textsc{S}(q)\, S_g+(1-f_\textsc{S}(q))\,S_l \label{fractionS}
\end{equation}
where $f_\textsc{S}(q)$ is the "fraction" of the bulk gas contributing to the background structure factor. The framework is completed by integration of $S^{bg}(z;q)$ over $[-L_g,L_l]$ determining that $S^{bg}(q)$ is of the form (\ref{Sback}), where $\zeta(q)$ is defined by analogy with the Gibbs dividing surface, Eq.~(\ref{ZG}):
\begin{equation}
\int\! dz\,\Big(\,S^{bg}(z;q)-S_g-(S_l-S_g)\,\theta\big(z-\zeta(q)\big)\Big)=0\,.
\end{equation}
The properties of $\zeta(q)$ and $\sigma_\textup{eff}(q)$, obtained via (\ref{sigmaeff}), are linked as seen in the rigidity coefficient, defined from the low-$q$ expansion $\sigma_\text{eff}(q)=\sigma+K_\textup{eff}\,q^2+\cdots$, which depends explicitly on $\zeta(0)$:
\begin{equation}\label{Rdty}
      K_\textup{eff} =\frac{5\sigma}{4}\left(\xi_g^2 -\xi_g\xi_l+\xi_l^2 -\frac{2}{5}Z_G\,\zeta(0)\right)
\end{equation}
Note that for the case of Ising symmetry ($\xi_l=\xi_g$), the rigidity is uniquely determined as $K_\textup{eff}=5\sigma \xi_l^2/4$ \cite{Parry2014}
since $\zeta(q)=0$. However, with asymmetry, even the {\it sign} of the rigidity may be altered depending on $\zeta(0)$ and the asymmetry ratio $\kappa_g/\kappa_l$.\\

We can now consider the merits and physical interpretation of different separation schemes by noting that fixing any one of $f_\textsc{G}(q)$, $f_\textsc{S}(q)$, or $\zeta(q)$ determines the other two. Consider setting $f_\textsc{G}(q)=1/2$, i.e.~$G^{bg}(0,0;q)\,=\,\frac{1}{2}(G_g+G_l)$. This gives  $\zeta(q)=\frac{3}{4}(\kappa_l(q)^{-1}\!\!-\!\kappa_g(q)^{-1})$. Alternatively, setting $f_\textsc{S}(q)=1/2$, i.e.~$S^{bg}(0;q)\,=\,\frac{1}{2}(S_g+S_l)$, leads to  $\zeta(q)=\frac{1}{2}(\kappa_l(q)^{-1}\!\!-\!\kappa_g(q)^{-1})$. These are not entirely implausible, but in both cases the respective "fractions" of bulk gas and liquid contributing to the background correlation function and structure factor are completely {\it ad hoc}. Thus, the choice $f_\textsc{G}(q)=1/2$ corresponds to $f_\textsc{S}=(3\kappa_g(q)+\kappa_l(q))/(4(\kappa_g(q)+\kappa_l(q)))$ which takes a markedly different value from $1/2$ when $\kappa_g\gg\kappa_l$. Similarly, $f_\textsc{S}(q)=1/2$ implies $f_\textsc{G}=\kappa_l(q)/(\kappa_g(q)+\kappa_l(q))$, which is very different from $1/2$ except for the case of Ising symmetry. The inconsistency can be avoided by requiring that $f_\textsc{S}(q)=f_\textsc{G}(q)$, which determines $f_\textsc{S}(q)=\kappa_g(q)/(\kappa_g(q)+\kappa_l(q))$. This is the {\it only} scheme for which the fractions of liquid and gas contributing to $S^{bg}(0;q)$ are identical to those for $G^{bg}(0,0;q)$ for {\it all} $q$, and leads to
\begin{equation}
\zeta(q)=\kappa_l(q)^{-1}-\kappa_g(q)^{-1}\;.
\label{zetatrue}
\end{equation}
Note that $\zeta(0)=Z_G$ so that the weighting of $S_g$ and $S_l$ in $S^{bg}(q)$ at $q = 0$ over the interval $[-L_g,L_l]$ is exactly the same as that of $\rho_g$ and $\rho_l$ in the total number of particles (per unit area) $N=\int\! dz\,\rho(z)=(L_g+Z_G)\rho_g+(L_l-Z_G)\rho_l$. However, $\zeta(q)$ vanishes as $q$ increases, which shifts the liquid-gas balance in (\ref{Sback}) to $z=0$, where the density gradient is maximum. The $q$ dependence of the weighting lengthscale, which is also present for the other two schemes, has not been appreciated previously. Note that going beyond the present mean-field treatment by allowing for capillary-wave induced broadening of the interface has a minor effect: the result for $Z_G$ is unchanged, and the position of the maximum in the density gradient is only weakly shifted \cite{SM}. Within the present DP calculation, the result $\zeta(0)=Z_G$ follows also when one notes that the condition $f_\textsc{S}(q)=f_\textsc{G}(q)$ implies that $S^{bg}(z;q)$ is continuous and differentiable at the origin. Thus, $S^{bg}(z;0)$ has exactly the same shape as the density profile. This also means that all other separation schemes lead to a $S^{bg}(z;q)$ which has a kink at the origin. Results for different choices of $\zeta(q)$ are shown in Fig.~1 of \cite{SM}. The effective tension follows as
\begin{equation}
\frac{\sigma_\textup{eff}(q)}{\sigma}=\frac{2\kappa_g(q)^2\kappa_l(q)^2}{\kappa_g\kappa_l(\kappa_g(q)+\kappa_l(q))}\left(\frac{1}{\kappa_g(q)\!+\!\kappa_g}+\frac{1}{\kappa_l(q)\!+\!\kappa_l}\right),
\label{sigmatrueasym}
\end{equation}
yielding a rigidity $K_\textup{eff}=\left(3\xi_g^2-\xi_g\xi_l+3\xi_l^2\right)\sigma/4$, which remains positive for all values of $\kappa_g/\kappa_l$.\\

\begin{figure}[t]
\includegraphics[width=\columnwidth]{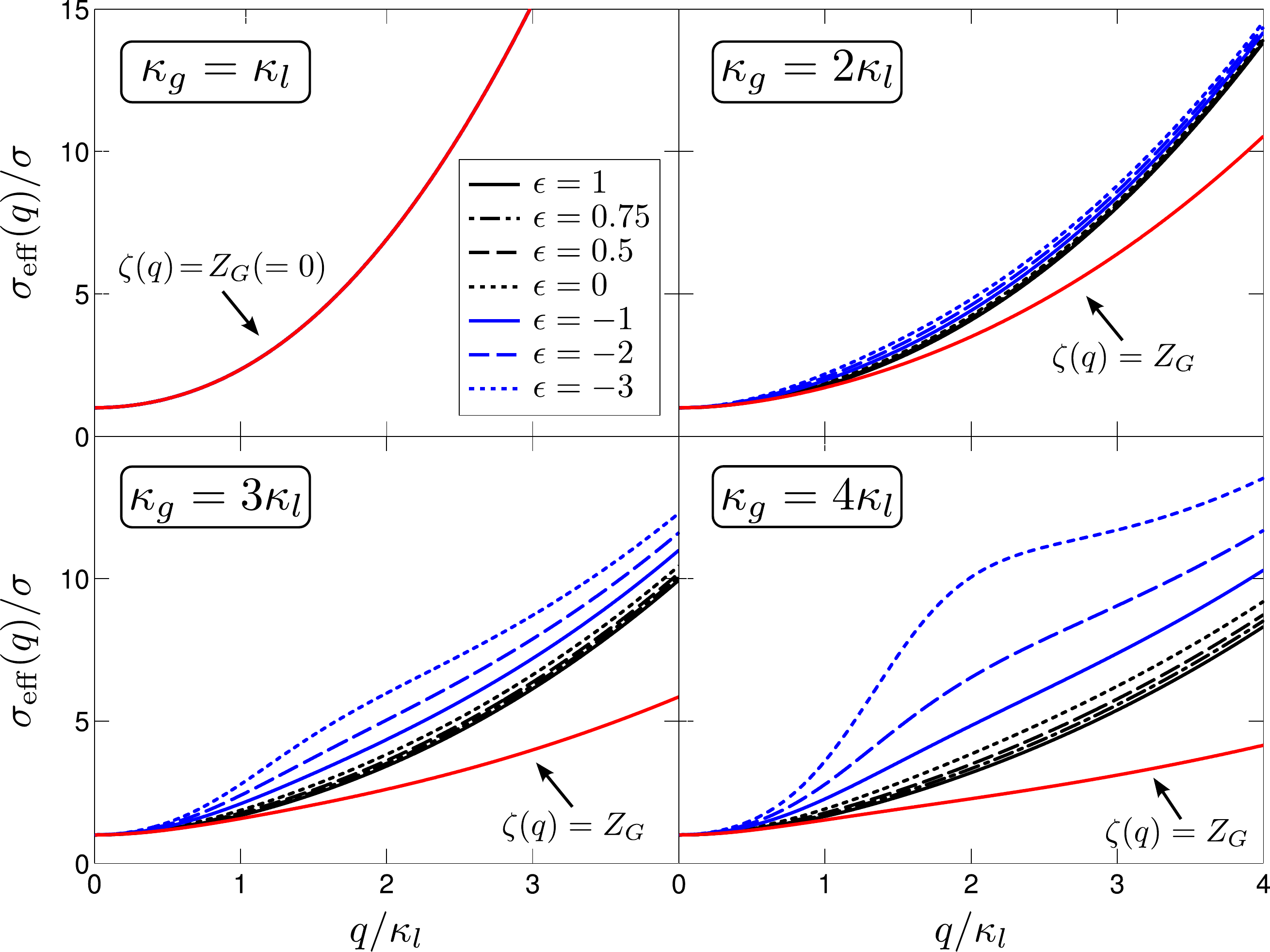}
\caption{\label{two}Variation of the effective surface tension $\sigma_\textup{eff}(q)$ for increasing liquid/gas asymmetry obtained for different choices of the microscopic background weighting lengthscale $\zeta(q)=\epsilon(\kappa_l(q)^{-1}\!-\!\kappa_g(q)^{-1})$, and for $\zeta(q)=Z_G$. In the case of Ising symmetry (top left) all results are identical. According to simple DFT estimates \cite{Lu1985}, $\kappa_g=2\kappa_l$ corresponds to  $T/T_c\approx 0.75$, and $\kappa_g=4\kappa_l$ to $T/T_c\approx 0.6$.}
\end{figure}

Next, consider schemes which instead set $\zeta(q)$ to be a ($q$-independent) constant. Consider, for example, setting $\zeta(q)=Z_G$. While this initially appears desirable, it requires that $S^{bg}(0;q)$ is {\it lower} than $S_g$ (even negative) for sufficiently large $q$ (see \cite{SM}). This arises as the reasonable requirement $S_g\le S^{bg}(0;q)\le S_l$ leads naturally to the bounds $-\kappa_g(q)^{-1}\le\zeta(q)\le \kappa_l(q)^{-1}$, implying that $\zeta(q)$ must vanish as $q$ increases. A final choice $\zeta(q)=0$ corresponds to the plane where $S(z;q)$ (and $\rho'(z)$) is largest. In this case, there is no problem with the behaviour of $S^{bg}(0;q)$, for which $f_\textsc{S}(q)=\kappa_l(q)/(\kappa_g(q)+\kappa_l(q))$. However, this choice is equivalent to setting $f_\textsc{G}(q)= (2\kappa_l(q)-\kappa_g(q))/(\kappa_g(q)+\kappa_l(q))$ in (\ref{fractionG}) which is {\it{negative}} for even moderate liquid/gas asymmetry. Thus, when $\kappa_g\gg\kappa_l$, the fraction of the bulk gas contributing to the background correlation function is $f_\textsc{G}(q)\approx -1$ while the fraction for the liquid is $1-f_\textsc{G}(q)\approx 2$. In Fig.~\ref{two}, we show plots of $\sigma_\textup{eff}(q)$ obtained for different separation schemes and for increasing asymmetries $\kappa_g/\kappa_l$. For the case of pure Ising symmetry ($\kappa_g=\kappa_l$), all are equivalent and $\zeta(q)=Z_G=0$. When $\kappa_g>\kappa_l$, the difference with (\ref{sigmatrueasym}) is always largest for the choice $\zeta(q)=Z_G$, which underestimates significantly $\sigma_\textup{eff}(q)$ at large $q$, even though it identifies the rigidity correctly. The opposite is true for the other separation schemes which work reasonably well at large $q$ but, alas, have different rigidities, consistent with Eq.~(\ref{Rdty}).\\

In summary, explicit DFT results for the simple DP model, tell us that while in principle there is freedom to choose any separation, $S(q)=S^{bg}(q)+S^{ex}(q)$, in all bar one case there was no physical interpretation of what the "background" means in terms of weighted bulk contributions. In the "physical" case, Eq.~(\ref{zetatrue}), the weighting lengthscale $\zeta(q)$ is $q$-dependent. Let us consider the wider implication. Suppose that for another model, simulation or experimental study we have two different separation schemes with different $\zeta(q)$ and $\sigma_\textup{eff}(q)$. Since the total structure factor $S(q)$ is the same in both descriptions, the difference in the inverse $q$-dependent surface tensions follows from (\ref{sigmaeff}) and (\ref{Sback}):
\begin{equation}
\Delta\left(\frac{1}{\sigma_\textup{eff}(q)}\right)\;=\;q^2\,\frac{S_l-S_g}{(\rho_l-\rho_g)^2}\;\Delta \zeta(q)
\label{uncertainty}
\end{equation}
where $\Delta \zeta(q)$ is the difference in weighting lengthscales. The two schemes agree the value of $\sigma_\textup{eff}(q)$ in the limit $q\to 0$ must be $\sigma$, but disagree at larger $q$, though the difference $\Delta \zeta(q)$ remains {\it microscopic}. The expression (\ref{uncertainty}) can therefore be viewed as characterising a fundamental {\it uncertainty} in the $q$-dependent surface tension arising from the indeterminacy of $\zeta(q)$. This would not be a problem if the weighting lengthscale $\zeta(q)$ was equal to the position of the Gibbs dividing surface for all $q$. However, the fact that $\zeta(q)$ is $q$-dependent, even in this very simple DFT, means that this must be the case more generally. Therefore, unless $\zeta(q)$ can be measured or determined {\it independently}, there will always be an uncertainty in the excess contribution to the structure factor implying that the behaviour of $\sigma_\textup{eff}(q)$ away from $q=0$ is essentially unknowable. In this regard, the limits of certainty on the form of $\sigma_\textup{eff}(q)$ arising from an unknown microscopic lengthscale are reminiscent of the high wavevector cutoff $\Lambda$ used in classical CW theory.
One arrives at very similar conclusions if one defines a $q$-dependent tension via the local structure factor, instead of (\ref{sigmaeff}). For example, one could measure where $S(z;q)$ is largest (in our case $z=0$),  and define,  $S^{ex}(0;q)=\Delta\rho\,\rho'(0)/q^2 \sigma_\textup{eff}(q)$ which generalises (\ref{CW2}). In this case, we are still left with an uncertainty in $\sigma_\textup{eff}(q)$, similar to (\ref{uncertainty}) except that $\Delta\zeta(q)$ is replaced by $(\xi_g+\xi_l)\Delta f_\textsc{S}(q)$, where $\Delta f_\textsc{S}(q)$ is the uncertainty in the $q$ dependence of the weighting fraction $f_\textsc{S}(q)$.\\ 

In our analysis, we found that $\zeta(0)=Z_G$; this result was not imposed. Rather, it emerged from trying to identify a consistent choice for the fractions $f_\textsc{S}(q)$ and $f_\textsc{G}(q)$ in the background $S^{bg}(0;q)$ and $G^{bg}(0,0;q)$. It would be extraordinary if, beyond the present DP model, a suitable separation of the local structure factor {\it always} results in a weighting lengthscale satisfying $\zeta(0)=Z_G$. One way of imposing this would be to set $S^{bg}(z;0)-S_g\propto \rho(z)-\rho_g$. Although this is valid for the separation leading to (\ref{zetatrue}) in the present DP model, it cannot be generally valid. Recall that, beyond mean-field, the profile is strongly affected by thermal wandering which cannot, by definition, be in the background contribution. This leaves us with two scenarios which determine the robustness of the $q$ expansion of $\sigma_\textup{eff}(q)$. First, if as found here, sensible separations always find $\zeta(0)=Z_G$, then the uncertainty $\Delta\zeta(q)$ vanishes as $q\to 0$. From (\ref{uncertainty}), this means that one may then write $\sigma_\textup{eff}(q)=\sigma + A\,q^2\ln q+K_\textup{eff}\,q^2+\cdots$ and identify a meaningful rigidity $K_\textup{eff}$. Nothing else, however, can be said unless the $q$ dependence of $\zeta(q)$ is determined. Second, if one finds instead that there are, in general, different equally acceptable ways of separating $S(z;q)$ into background and excess terms which sometimes result in $\zeta(0)\ne Z_G$, then the uncertainty $\Delta\zeta(0)\ne 0$ implies that only the term $\mathcal{O}(q^2\ln q)$ induced by the dispersion forces remains well characterised. In this second scenario, which appears more likely to us, it is not just that $K_\text{eff}$ is non-unique but that the separation of $S(q)$ into background and excess terms may well be ill-defined.\\


We thank Edgar Blokhuis, Pedro Tarazona, Enrique Chac\'on, Felix H\"ofling and Gary Willis for extremely illuminating correspondence and discussions. AOP acknowledges the EPSRC, UK for grant EP/J009636/1. CR acknowledges support from Ministerio de Econom\'{i}a y Competitividad (Spain) Grant FIS2010-22047-C05.
%

\bibliography{wetting}

\end{document}


\title{Supplementary Information for "\textit{Liquid-Gas Asymmetry and the Wavevector-Dependent Surface Tension}"}

\author{A.O.\ Parry}
\affiliation{Department of Mathematics, Imperial College London, London SW7 2BZ, UK}

\author{C.\ Rasc\'{o}n}
\affiliation{GISC, Departamento de Matem\'aticas, Universidad Carlos III de Madrid, 28911 Legan\'es, Madrid, Spain}

\author{R.\ Evans}
\affiliation{HH Wills Physics Laboratory, University of Bristol, Bristol BS8 1TL, United Kingdom }

\maketitle

Here we provide further details of three calculations: The determination of $G(z,z';q)$, its separation into background and excess terms, and the influence of capillary-waves fluctuations on lengthscales characterising the interfacial profile. A plot showing the background local structure factor $S^{bg}(z;q)$ for different choices of $\zeta(q)$, and the location of the Gibbs dividing surface $Z_G$ is also included.

\section{The expression for $\;G(z,z';q)$}

$G(z,z';q)$ is obtained from solving the Ornstein-Zernike equation which, for our local square-gradient functional, reduces to \cite{Evans1981}
\begin{equation}
\left(-\frac{{\partial}^2}{\partial z^2}+q^ 2 +\phi''(\rho(z))\right)G(z,z';q)=\delta(z-z')
\label{OZ}
\end{equation}
where, as earlier, we have set $k_B T=1$ and $b=1$. Far from the interfacial region, the correlation function tends to the appropriate bulk expression:
\begin{equation}
G_a(|z-z'|;q)\;=\;\frac{e^{-\kappa_a(q)\,|z-z'|}}{2\,\kappa_a(q)}
\label{Ggaslig}
\end{equation}
on the gas ($a=g$) and liquid sides ($a=l$) of the interface, respectively. Within the DP model, the second derivative of the potential can be written $\phi''(\rho(z))=\kappa_g^2 +(\kappa_l^2-\kappa_g^2)\,\theta(z)-(\kappa_g+\kappa_l)\,\delta(z-z')$, and the Ornstein-Zernike equation is readily solved using elementary methods. The correlation function takes its maximum value when both particles are at the origin (which have located at the point where the derivative $\rho'(z)$ is maximal), and is given by
\begin{equation}
G(0,0;q)=\frac{1}{q^2}\,\frac{1}{\,(\kappa_g(q)+\kappa_g)^{-1}+(\kappa_l(q)+\kappa_l)^{-1}}
\label{G002}
\end{equation}
Thus, the correlation function exhibits the anticipated Goldstone mode singularity as $q \to 0$. Then, if the particles are on the same side of the interface ($zz'>0$), $G(z,z';q)$ is given by
\begin{equation}
G=G_a(|z-z'|;q)+(G(0,0;q)-G_a(0;q)) e^{-\kappa_a(q)(|z+z'|)}
\label{Gsameside}
\end{equation}
with $a=g$ or $a=l$, as appropriate. However, if these lie on either side (say, $z<0<z'$),
\begin{equation}
G\;=\;G(0,0;q)\,e^{-\kappa_g(q)|z|}\;e^{-\kappa_l(q)z'}
\label{Gasym}
\end{equation}
The correlation function is a continuous function of $z$ and $z'$.

\begin{figure}[t]
\includegraphics[width=\columnwidth]{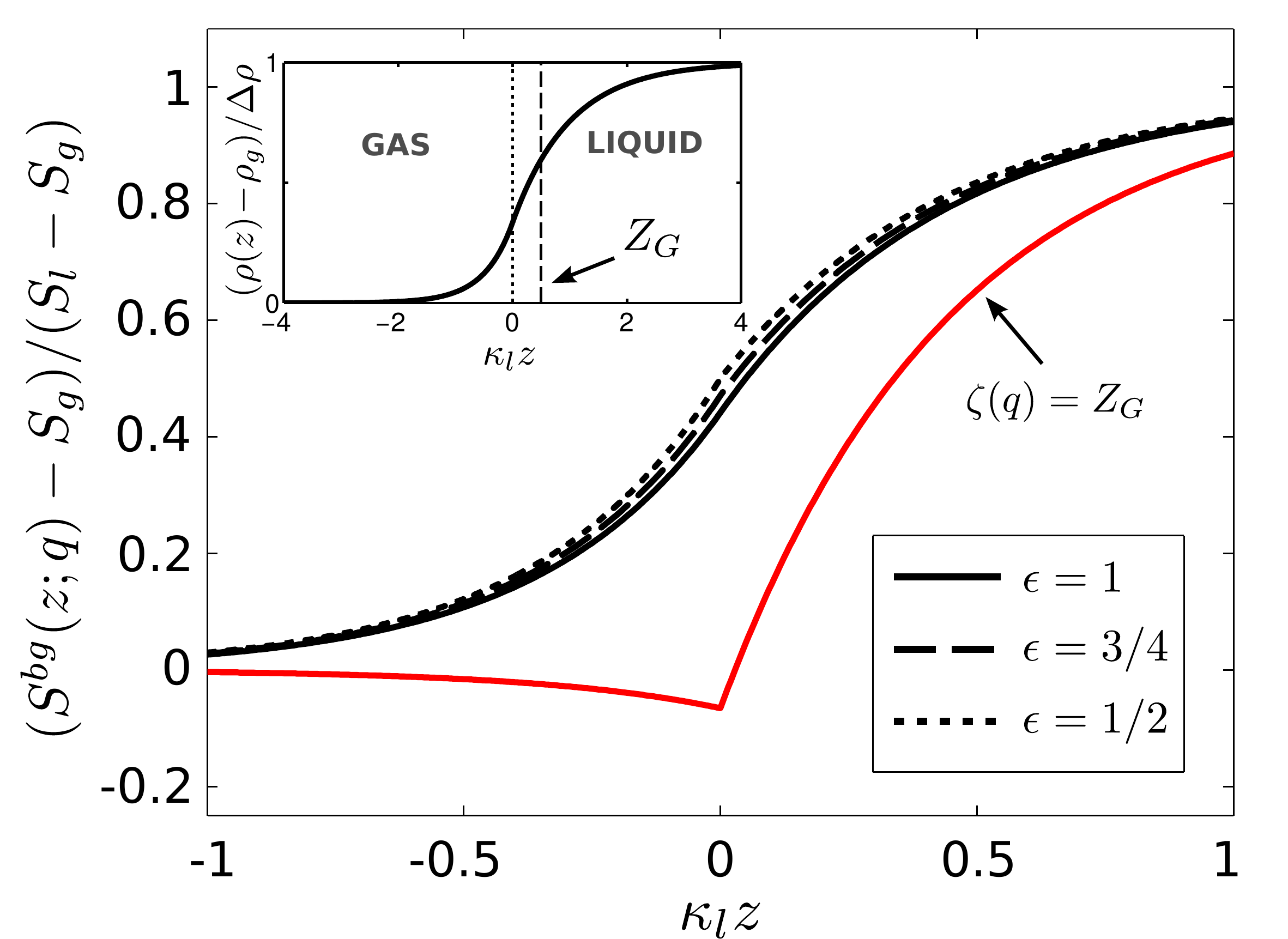}
\caption{\label{one}Background local structure factor for three different choices of the microscopic weighting lengthscale $\zeta(q)=\epsilon(\kappa_l(q)^{-1}\!-\!\kappa_g(q)^{-1})$, and for $\zeta(q)=Z_G$ (red line). Here, $\kappa_g/\kappa_l=2$, and $q=2\kappa_l$. For  $\epsilon\ne 1$, corresponding to schemes where  $f_\textsc{S}\ne f_\textsc{G}$, the $S^{bg}(q;z)$ has a kink at the origin. For the scheme with $\zeta(q)=Z_G$, $S^{bg}(z;q)$ is lower than $S_g$ for $z\le 0$. The inset shows the density profile and the location of the Gibbs dividing surface, $Z_G$.}
\end{figure}

\section{The separation of $G(z,z';q)$}

\subsection{Ising symmetry: $\kappa_l=\kappa_g$}

The expression for $G(z,z'q)$ simplifies considerably when $\kappa_l=\kappa_g\equiv \kappa$ (i.e. Ising-like symmetry). The bulk liquid and gas phases have identical correlation functions and structure factors which we write as
\begin{equation}
G^b(|z-z'|;q)\;=\frac{e^{-\kappa(q)|z-z'|}}{2\kappa(q)}, \hspace{0.5cm}S^b(q)=\frac{1}{\kappa^2+q^2}
\label{Gliquid}
\end{equation}
where $\kappa(q)=\sqrt{\kappa^2+q^2}$. In this case, the position dependence of the correlation function throughout the inhomogeneous region can be expressed as \cite{Parry2014}
\begin{equation}
G(z,z';q)\,=\,G^b (|z-z'|;q)\,+\,\frac{G^b(|z|;q)\,G^b(|z'|;q)}{G^b(0;0)-G^b(0;q)}
\label{G1}
\end{equation}
This result splits {\it unambigously} into a background term, equal to the bulk correlation function, and an excess part
\begin{equation}
G^{ex}(z,z';q)=\frac{G^b(|z|;q) G^b(|z'|;q)}{G^b(0;0)-G^b(0;q)}, 
\label{GexIsing}
\end{equation}
which decays exponentially, as each particle moves away from the interface, and is controlled by the inverse lengthscale $\kappa(q)$. For $q\ll\kappa$, and distances close to the interface (\ref{GexIsing}) is entirely in keeping with the prediction of capillary-wave theory (Eq.~(1) in the main article). It follows that, for this Ising symmetric case, the local structure factor also separates \cite{Parry2014}
\begin{equation} 
S(z;q)\;=\,S^b (q)\,+\,\frac{S^b(q) G^b(|z|;q)}{G^b(0;0)-G^b(0;q)}
\label{S1}
\end{equation}
so that the background contribution is simply equal to the bulk $S_b(q)$. Similar to the correlation function, the excess contribution decays exponentially as the particle position $z$ moves away from the interface. Integration of $S(z;q)$ over the macrosopic interval $[-L_g,L_l]$ gives
\begin{equation}
S_{tot}(q)=(L_g+L_l)\, S^b(q)+\frac{S^b(q)^2}{\,G^b(0;0)-G^b(0;q)}
\end{equation}
where the first term is the total background contribution. From Eq.~(4) of the main article, it follows that $\zeta(q)=0$, and one may then identify a wave-vector dependent tension from the excess contribution as
\begin{equation}
\sigma_\textup{eff}(q)\;=\;\sigma\;\frac{2\,(1+q^2\xi_b^2)^2}{1+q^2\xi_b^2+\sqrt{1+q^2\xi_b^2}}
\label{exp}
\end{equation}
where $\xi_b=1/\kappa$ is the bulk correlation length. This is the result given in Eq.~(28) of \cite{Parry2014}.

\subsection{Liquid-gas asymmetry: $\kappa_l\ne\kappa_g$}

We wish to see if such a simple separation of $G(z,z;q)$ occurs also when there is a more general asymmetry between the bulk phases. That is, we wish to write
\begin{equation}
G(z,z';q)= G^{bg}(z,z;'q)+G^{ex}(z,z;'q)
\end{equation}
where now the background ($bg$) contribution necessarily involves a mix of bulk liquid and gas correlation functions. It is only permissible to construct the background and excess contributions using the same elementary exponential functions $e^{\pm\kappa_a(q)z}$ and $e^{\pm\kappa_a(q)z'}$ appearing in the equilibrium solution (\ref{Gsameside}) and (\ref{Gasym}). Otherwise, one is introducing entirely arbitrary and non-physical lengthscales which are neither incorporated in the bulk nor in the inhomogeneous system. It is also necessary that the background and excess contributions are continuous functions of $z$ and $z'$; otherwise, there is no connection with the standard capillary-wave expression (Eq.~(1) in the main article) as $q\to 0$ nor with effective Hamiltonian theory. Finally, the background must approach the appropriate bulk function far from the interface. This leaves us only with a background contribution $G^{bg}(z,z';q)$ of the form
\begin{equation}
G^{bg}=G_a(|z-z'|;q)+(G^{bg}(0,0;q)-G_a(0;q))\, e^{-\kappa_a(q)(|z|+|z'|)}
\end{equation}
if $zz'>0$ and 
\begin{equation}
G^{bg}=G^{bg}(0,0;q)\,e^{-\kappa_g(q)|z|}e^{-\kappa_l(q)z'}
\end{equation}
if $z<0<z'$. As for the symmetric Ising case (see (\ref{GexIsing})), the excess term is a simple product
\begin{equation}
G^{ex}(z,z';q) =G^{ex}(0,0;q)\;e^{-\kappa_a(q)|z|}\,e^{-\kappa_a(q)|z'|}
\end{equation}
This decomposition reproduces identically the equilibrium function $G(z,z;q)$, provided that we impose $G(0,0;q)= G^{bg}(0,0;q)+G^{ex}(0,0;q)$. This leaves us with only a single function characterising the whole separation: the value of $G^{bg}(0,0;q)$. Integration leads directly to the separation of the local structure factor described in the main article and to the consistency requirement $S^{bg}(0;q)=G^{bg}(0,0;q)(\kappa_g(q)^{-1}+\kappa_l(q)^{-1})$. Thus, one may equally consider $S^{bg}(0;q)$ as the function to characterise the separation; see Eqs.~(12) and (13) of the main article.\\

We emphasize that any separation/decomposition of the local structure factor $S(z;q)$ into background and excess, which dictates the separation of the total structure factor $S(q)$ as defined in the main article, \textit{must be consistent} with the corresponding separation of $G(z,z';q)$. It is the latter, the density-density correlation function (clearly, the fundamental statistical mechanical quantity) which describes the structure of the interface at the two-particle level.\\

\section{The influence of capillary-wave fluctuations on the density profile}

Beyond mean-field approximation, the equilibrium profile $\rho_{eq}(z)$ is broadened by interfacial fluctuations. This is well-described by the approximate capillary-wave formula \cite{Weeks1977}
\begin{equation}
\rho_{eq}(z)\approx\int\!d\ell\;\, \rho_0(z-\ell)\, P(\ell)
\end{equation}
where here $\rho_0(z)$ may be taken to be the underlying mean-field profile (Eq.~(7) in the main article) and $P(\ell)$ is the probability distribution for the interfacial position $\ell$. If the interface has finite transverse area $L_\parallel^2$, and there are no other pinning fields, then $P(\ell)$ has a simple Gaussian form $P(\ell)=e^{-\ell^2/2\xi_\perp^2}/\sqrt{2\pi}\,\xi_\perp$, where the interfacial width $\xi_\perp$ satisfies the celebrated capillary-wave formula $\xi_\perp^2=(2\pi\sigma)^{-1}\ln (L_\parallel\Lambda)$ (recall that $\sigma$ is the surface tension, and $\Lambda$ the cut-off). In the thermodynamic limit, $L_\parallel\to\infty$, the equilibrium profile is very different from the underlying mean-field profile, and is well-described an error function of width $\xi_\perp$ \cite{Weeks1977}. However, the position of the Gibbs dividing surface $Z_G$ and the location of the maximum in the density gradient $Z_0$ are hardly affected. In fact, the normalization condition on $P(\ell)$ means that the location of $Z_G$ is completely unaltered from its mean-field prediction, $Z_G=\xi_l-\xi_g$. Similarly, asymptotic analysis of the integral shows that the maximum in the density profile (which occurs at $Z_0=0$ in mean-field approximation) now occurs at
\begin{equation}
Z_0\;\approx\;-3\;\frac{Z_G(\xi_g^2+\xi_l^2)}{\xi_\perp^2}
\end{equation}
and, therefore, remains close to the origin as $L_\parallel\to\infty$, similar to the mean-field prediction.

\bibliography{wetting}